# Human Computations in Citizen Crowds: A Knowledge Management Solution Framework

NADEEM KAFI*, ZUBAIR AHMED SHAIKH*, AND MUHAMMAD SHAHID SHAIKH**

## ABSTRACT

KG (Knowledge Generation) and understanding has traditionally been a Human-centric activity. KE (Knowledge Engineering) and KM (Knowledge Management) have tried to augment human knowledge on two separate planes: the first deals with machine interpretation of knowledge while the later explores interactions in human networks for KG and understanding. However, both remain computer-centric. Crowdsourced HC (Human Computations) have recently utilized human cognition and memory to generate diverse knowledge streams on specific tasks, which are mostly easy for humans to solve but remain challenging for machine algorithms. Literature shows little work on KM frameworks for citizen crowds, which gather input from the diverse category of Humans, organize that knowledge with respect to tasks and knowledge categories and recreate new knowledge as a computer-centric activity.

In this paper, we present an attempt to create a framework by implementing a simple solution, called ExamCheck, to focus on the generation of knowledge, feedback on that knowledge and recording the results of that knowledge in academic settings. Our solution, based on HC, shows that a structured KM framework can address a complex problem in a context that is important for participants themselves.

## 1.    INTRODUCTION

Civic innovations and community problem solving require new dimensions in crowdsourcing and HC to overcome two important shortcomings: (1) a knowledge-centric view of human workers based on their diverse capabilities and skills as communities, and (2) expressing human-centric knowledge creation and management as workflow of linked HC tasks. These shortcomings, when addressed, not only extend existing body of research but also deliver direct benefits, at large scale, to governments, organizations and institutions alike. Innovations utilizing citizen science projects, citizen-centric governance, and customer-centric business practices are the next frontiers of digital economy utilizing human workers for knowledge worker.

Authors E-Mail: (nadeem.kafi@nu.edu.pk, zubair.shaikh@jinnah.edu, shahid.shaikh@sse.habib.edu.pk)
*       Department of Computer Science, FAST-National University of Computer & Emerging Sciences, Karachi.
* *     Department of Electrical Engineering, Habib University, Karachi, Pakistan

The social era, driven by ubiquitous and social computing, has provided unconceivable access to the human networks, forming communities of practice managing different type of knowledge for different innovations and problem solving [1]. Crowdsourcing "can [also] take the form of peer-production (when the job is performed collaboratively)" [2], utilize collective knowledge of workers as collective intelligence to execute complex and creative tasks. This community-based paradigm, part of the pivotal definition, not only motivate workers to innovate but also enrich collective knowledge of the community with skills, experiences, digital artifacts and social sensing data. HC [3] are part of a computational process, which strategically directs humans to perform tasks using cognitive skills and provide data. Crowdsourced HC enable "use of humans to perform some form of computation…that leverage human intelligence but challenges even the most sophisticated AI (Artificial Intelligence) algorithms that exist today" [4]. We generalize a neutral term "citizen crowds" for all type of human workers, focusing on their diverse contributions and disregarding any affiliation to organizations and institutions. We assume that citizen crowds in communities become knowledge workers [5,6].

KM in citizen crowds makes use of inbuilt human habits of participatory problem-solving to externalize human cognition and memory, including machine-centric argumentation through information management and strategy encapsulated in process workflows [7]. In general, communities seek innovative outcomes when a large number of their members face similar problems, which we call PCI (Problems of Common Interest). These problems mainly relate to people such that they face and solve them using collaboration and collective intelligence. In terms of crowdsourced HC, both requestors and workers share problem and solution domains. We believe that the core underlying problem is in selecting one out of many solutions, thus the need to expressed the problem as a KM process.

Humanizing and socializing AI has become an important research problem, as AI would augment and "advance human behavior in complex human situations". The time has come "for AI to interact with humans, about human issues, to promote human growth" [8]. Therefore, a pertinent question is; how innovative civic services and community problems of future smart cities will be solved by hybrid human-machine systems, in real-time and using citizen crowds. Another relevant example, by Branson et. al. [9] reports a human-machine visual recognition system. This system incorporates different computer vision algorithms into a human-in-the-loop framework, where human and machine work to maximize each other's strengths. We believe that human interactions and involvement in AI have applications beyond personal assistants and navigational applications, but into applications where it intermingles with everyday people lives, such as in civic innovation sand community problem solving. However, the challenge of creating and managing knowledge in citizen crowds cannot take shape with rating and liking contents, as these "tools do not engage a human's perceiving, reflecting, meaning-making and deciding in the absence of absolutes" [8]. This defines the extended role that citizen crowds of future smart cities will undertake i.e. providing knowledge creation and management as a civic duty. Local governments and businesses, in reciprocity, shall be obliged to extend a citizen-centric model of operations and governance, where functional elements are driven by the knowledge creation ability of the citizen crowds.

The diverse functional requirements for human cognitive inputs for civic innovation and community problem solving require HC where workers have the right context and are co-located in the problem domain. Citizen crowds are ideal in this case, but the participatory and volunteer nature of input is difficult to obtain from existing tasks based crowdsourcing platforms. Citizens are direct beneficiaries whenever innovative solutions are implemented for civic services or community problems. First, it gives impetus for participation (social cause) but also provides the required knowledge and sensor data in the problem domain and human-centric solutions in hybrid human-machine systems. Second, identification and generation of new tasks, as a result of human input, are automated through machine algorithms.

The need for a framework that makes sophisticated hybrid human-machine computations possible is paramount, such that availability of human cognition and memory becomes on-demand, scalable, and distributed but location transparent similar to present day cloud computing. This model will allow much tighter integration of human processing with machine algorithms and extend beyond the present notion of "human in the loop". We argue that a KM model customizable to citizen crowds solves many issues hence the lack of a citizen-centric KM framework. This hinder government and organizations to successfully integrate people contributions, cognitive and human input data, in governance and monitoring systems and losses innovative inputs when crowds form complex opinions and provide their knowledge in both problem and solution domains. In this paper, we attempt to propose a citizen crowd based KM system where workers are part of both problem and solution spaces. They have "stake" in the problem space and as a result, take "ownership" of the problem-solving process by becoming part of the solution through their contributions. However, to the best of our knowledge, we have not found any literature reference of crowdsourced HC for KM at the time of this writing.

This paper has two contributions: First, it describes ExamCheck; a web-based application to study the creation of an interaction space using communities and crowdsourced HC for knowledge creation, sharing and use. ExamCheck implements a scenario that aligns teaching and assessments with CLOs (Course Learning Objectives). Second, based on the findings of ExamCheck, we extend traditional KM into crowd-based KM by outlining a structured KM framework based on HC in citizen crowds. This framework describes a complex problem as a workflow of HC tasks, creates human interactions spaces based on citizen crowds, execute the KM workflow using the human workers, and process the resulting human contributions to create diverse knowledge streams utilizing hybrid human-machine computations.

This paper is organized in seven sections: this introduction, background and related work, the scanrio-alignment of learning objectives, implementation and experimental setup, findings and challenges, our KM framework, conclusions and future work.

## 2. BACKGROUND AND RELATED WORK

The definition of knowledge is difficult to frame. In our context, knowledge takes one of two forms either as a resource or as a process. Knowledge as an "input resource" can be transformed using "process knowledge" to generate new knowledge as an "output resource" [10]. However, in our view of KM in citizen crowds should use Demarest [11] notion of commercial knowledge instead of philosophical or scientific knowledge. Commercial knowledge is social and transacted between individuals and organization in various forms.

KM aids innovation in organizations and networked communities of practice by addressing their specific structural characteristics, economics goals, ways of applying technology and behavioral aspects of people being involved [12]. The CSCW community view on knowledge and expertise sharing has also been shifted from object-centric to people-centric, and is grounded in social context of knowledge work practices and communication between knowledgeable humans [13]. KM poses two challenges: First, knowledge creation through the conversion of un-digitized tacit knowledge into explicit knowledge by interactions and transformations performed in people spaces [14,15]. Second, it specify a process of sharing, managing storage/retrieval and use of knowledge assets in terms of people and artifacts [16]. KM systems provide four interdependent knowledge processes: (1) creation, (2) storage and retrieval, (3) transfer, and (4) application to mitigate the challenges posed by KM. A variety of KM approaches and different KMS are reported in the literature due to "the diversity of knowledge types and attributes" [17]. We argue that KM is a knowledge-aware activism for managing progress including problem-solving. It has goal-based activities where individuals in a human network search, share, use and create tangible and non-tangible artifacts of knowledge within the boundaries of a mutually agreeable solution space.

Innovation and problem solving are fundamentally human-centric and process oriented, and are best described in terms of workflows. They capture the semantics of a KM processes and the human sociability in term of a collaboration architecture where humans act as "components and connectors", in order to execute them. Workflows are reusable elsewhere [18,19]. Workflows relate human "work activities, their resources, and communications between the actors", as part of a goal oriented KM, and keep track of history knowledge work [20]. In this paper, we have captured individual human contributions as HC tasks (both HC tasks and other tasks like sharing, search tasks, etc.) and then describe a solution in terms of a workflow of HC tasks we called "KM workflow". According to Assudani [10], these workflows contains "process knowledge" and are themselves subject to KM

Knowledge applied for the benefits of communities is not created in isolation but a shared space and context for emerging relationships that promote meaning. The concept of Ba (Nonaka and Konno [14]) defines and attaches knowledge with interactions perform in close proximity inside a physical or virtual space, resulting in the exchange of information (face-to-face, verbal, written). Ba also facilitate queries and clarifications in order to underst and the context, situation, and the interpretative mindset of the information provider. Therefore, both experiences of the receiver and those of the provider(s) are important. This emphasizes that information and data if interpreted in isolation with the provider interpretative framework, context, situation, time and the intended problem domain might result in invalid acquired knowledge by the receiver. Knowledge acquired in Ba (learning space) is intangible, and if seen in isolation, becomes tangible information and then freely communicated, stored and processed. Citizen crowds seeking innovative problem-solving create a Ba like space for interactions. Hence execution of KM workflow tasks place fulfilling of the requirements.

The execution of KM workflow in citizen crowds creates new knowledge by overcoming the KM challenges by implementing four interdepending KM processes of (1) creation, (2) storage and retrieval, (3) transfer, and (4) application. The key to knowledge creation is a dynamic process of dialogue and practice and referred it to as dialectic thinking and acting, in which contradictions are synthesized and resolves between people, organization, and environment [21]. Similarly, Thomas et al. [22] introduced "knowledge community: a place within which people discover, use, and manipulate knowledge, and can encounter and interact with others who are doing likewise" and suggested that social computing and knowledge socialization for building Knowledge Communities. Crowd-based knowledge creation is facilitated by social situations and sources thatprovide the necessary influence and help them learn while they construct knowledge using their own cognitive skills and understandings [23].

Knowledge creation, therefore, requires repetitive human interactions to understand and refine the context, sharing information artifacts, their assessment, and case based aggregation of contributions. As Yu and Nickerson [24], explained that crowds as "collective minds" exhibiting cobbler like behavior by taking work of others, and mend it to produce less error-prone work in an iterative way producing innovative outcomes. Human verification of others' work in each iteration also ensures the quality of outputs as a side effect. They also presented crowd-based participatory "an Internet-scale idea generation system" where evolutionary computation is applied to crowds [25]. This shows that human spaces are ideally suited for combining machine and HC together for knowledge contribution in peer productions.

Knowledge creation depends on the entity doing the interpretation: machine interpretable knowledge is the discipline of KE, and human interpretation of knowledge is the purview of KM [26]. AI has made constant efforts to model human intelligence; on the other hand, HC pursue solving tasks that are computationally challenging. These two research areas "hinge on the long-held belief that there exist problems that require human-level intelligence and reasoning to solve" [27]. When machine interpretations are not feasible or dependent on human interpretations then KM becomes the only source of knowledge creation.

Implementation of a HC systems to support a collaborative human architecture has many challenges. The inclusion of human in the computational process need considerations due to whims of human behavior that makes HC systems different from the deterministic machine computations. The "US Research Roadmap for HC" identified them as (1) participation, (2) analysis, (3) architecture, (4) design methods, and (5) infrastructure [28]. Reeves and Sherwood [29] also outlined five key consideration for designing systems with HC: (1)

designing for motivation and sustainability, (2) balancing system design and user practice through orientation and framing, (3) using situated-ness as a resource, (4) organizing human-machine relations, and (5) reconsider the utility of machine analogies in HC.

Many research efforts relating to hybrid human-machine computation have been reported in the literature. SmartSociety supports computing with hybrid human/ machine collectives, however, it comes short of identifying a KM model [30]. Retelny et. al. [31] experimentally shown that when teams among crowd workers are engaged, complex work can be performed using collective intelligence as a result of combined cognition, memory, and experience of the team members. They "envision a future of crowdsourcing with dynamic collaborations of diverse and interdependent participants". Brambilla et. al. [32] presented CrowdSearcher, a community-based crowdsourcing platform, where (complex) tasks are assigned to communities instead of individuals. Community control is achieved through rule-based adaptations by observing worker behaviors and their feedback. This work strengthens our community formation argument for complex task execution. However, the paper says nothing about community pre-selection for a given task, which is important in our agenda of crowd-based KM.

**Summary**: Civic innovation and community problem solving present unique challenge to citizen crowds where they become part of the solution when faced with problem of common interest. In this paper, we see the problem through the lens of KM in citizen crowds using human computations, and present an attempt to outline a KM framework based on implementing a simpler scenario. The new framework provides three core facilities: First, an innovation-seeking component where humans address innovation in terms of asking different questions (e.g. what they want, why they want it, how they want it).

Second, a knowing component that answers: who knows, what, where is what, etc. Third, an interaction space component where citizen crowds as knowledge workers contribute knowledge artifacts resulting in innovative outcomes. Finally, an aggregation and summarization component, which delivers customize knowledge to users involve in knowledge intensive activities.

# 3. THE SCENARIO-ALIGNMENT OF LEARNING OBJECTIVES

In order to concisely but effectively explain our point of knowledge creation and management in community crowds, we consider the problem of "constructive alignment". John Biggs [33] proposed "aligning teaching and assessing to course objectives by asserting that "students construct meaning through relevant learning activities" and the teacher does the alignment by setting an appropriate learning environment to support learning activities. We assume each course assessment as a written document, which is converted and/or stored as a digital artifact, and link knowledge socialization with assessment of artifacts that tell a "story", which the assessor interprets separately and later influence the grading community's opinions through aggregation [22]. This shared understanding helps graders to under answers w.r.t learning objectives, and identify misalignments with proposed suggestions/solutions for realign teaching with course learning objectives. We categorized our assessment activity "as the social system which shapes activity, provides a useful basis for elaborating the nature, organization and goals of the particular situation in which activity is undertaken" [23].

We conceptualize this alignment as a KM process (Fig. 1). This semester wide process, repeated for each course assessments (e.g. quizzes, assignments, midterm, final exam) measures student progress in various learning activities. These activities are based on CLOs set forth in the course outline and weekly plan. Students are engaged in assessments by self-reporting invalid or partially correct answers, peer grading, and feedback to graders. This post-classroom process also encourages them to share their reviews/feedback with other students and graders using a virtual interaction space that exists for the whole session. This KM process enables the teacher to adjust learning activities to achieve an optimal alignment with CLOs.

We have designed a web application to implement our KM activity that assesses answer books as its core task, called ExamCheck. In our scenario, the teacher is responsible for defining CLOs based on a course outline. These will form the basis for all assessments in the course. Students registered in the course form one community and we further register volunteers from two communities-Faculty and Alumni, to assess their answer books. However, we assume that volunteers can perform more knowledge-centric activities related to "constructive alignment", as a side effect and after being influenced by assessing the answer books.

After taking the exam, students snap their answer books in the presence of the TAs (Teaching Assistants) and uploads them to our web application. In a separate but associated task, students tag each assessment question with topic from a list of CLOs and provide a narrative as feedback (self-review). ExamCheck allows the teacher to select how different questions from each assessment are crowdsourced to faculty and alumni e.g. different graders can check the same question in order to increase transparency and reduce error in checking answer books. Registered faculty and alumni are invited to grade the assessment items, which they can either decline or forward to other un-registered acquaintances – possibly increasing the volunteers in their community. ExamCheck informs the teacher when all answer books are graded and s/he can release the results based on the maximum, minimum or average scores as one question will possibly check by different graders. ExamCheck interact with students again to take narrative feedback on graded questions, identifying gaps in their understanding of assessment items with that of graders. In the last iteration, teacher, faculty and alumni complete the KM activity by identifying any lesson learned for alignment and innovating contents and context of learning activity. Tasks performed within three different communities: students, faculty and alumni as shown in Table 1.

The ExamCheck scenario benefits the faculty in terms of reduced answer sheet checking effort/time, it also provides an interaction space, as a weak community of practice, where contributions from faculty, alumni and students are organized to create knowledge for learning objective centric alignment opportunities in the form of a lesson learned.The scenario shows that crowdsourced HC are a possible candidate for implementing computer-centric KM activities, especially in community crowds; however, the overlaps between this two disciplines and their key linkages need further investigation.

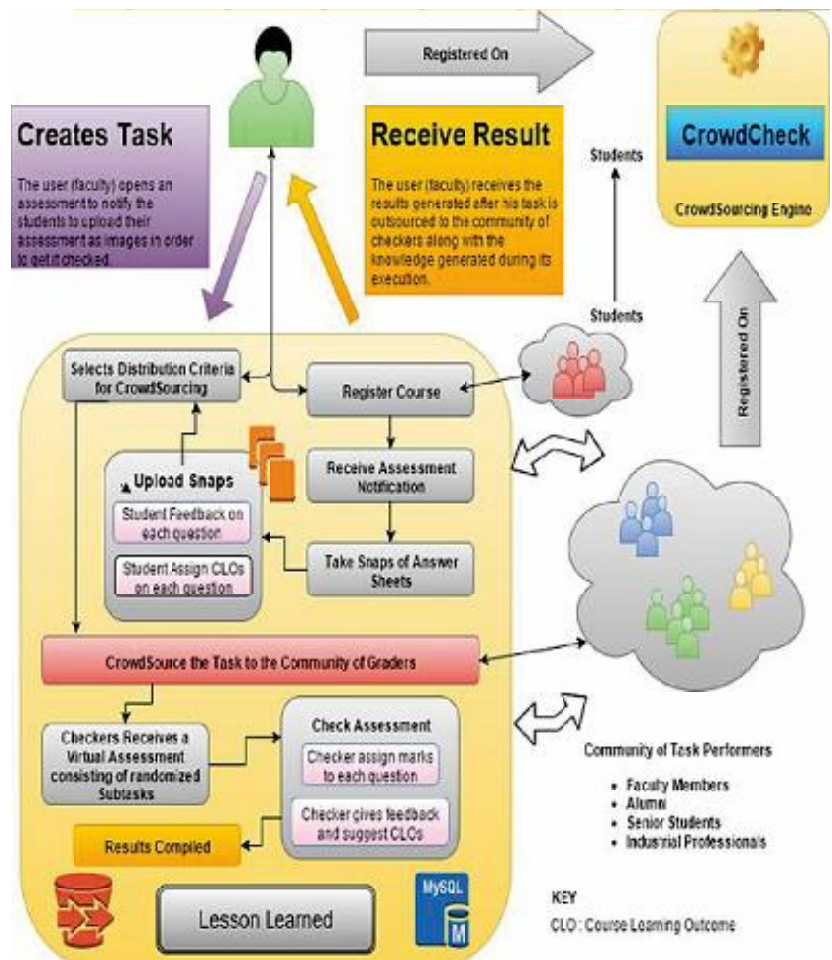

*FIG. 1. CONSTRUCTIVE ALIGNMENT SCENARIO*

## 3.1 Analysis

We now briefly analyze our scenario to identify requirements, affix the need for crowdsourcing, deliberatethe limitations and proposal solutions of existing crowdsourcing platforms.

**Requirements:** In our scenario, different questions are graded, on a predefined scale, against one or more CLOs. Additionally, the graders have access to students' pre-grading feedback, a sample solution of the question, and other graders using chat. This requires a people-centric knowledge creation approach,as graders utilize their cognitive skills and memory to take grading decisions, which are highly influenced by practices that are carefully followed and known within the grading community. These communities not only have intimate knowledge of the grading process, but also know the impact of grading process on the learning system and its stakeholders. In summary, we have two main requirements: (R1) an automated system for evaluation of handwritten artifacts (answer books) against pre-set objectives, allowing artifact creators (students) and evaluators (graders) as participants in a virtual interaction space. This allows capturing of semantic about the level of learning in a course on a per assessment basis. (R2) The need for a democratic, scalable, and transparent (open) knowledge creation mechanism,using the interaction space created around grading of assessments. The contributions of students and graders will help course teacher and senior faculty to agree on a new procedure, on per assessment basis, to align teaching and assessment with CLOs in between two assessments. (R3) A method to train incoming new graders about the underlying assessment process grounded in practices followed within the community.

**Need for Crowdsourcing:** The requirements for existing undergraduate courses have largely been multifaceted. Normally, expectations from a course exceed its offering, as it is difficult to include all possible contents from both academic and industry perspectives. Only large participation from all stakeholders will decide the right mix of learning objectives. Crowdsourced participation not only provides the required scalability (satisfying R1) but also made available human computations as an additional functionality for human-centric decision making in the context of the KM activityin our assessment scenario (required for R2). However, the complexity of our KM activity poses limitations for conventional crowdsourcing methods and platforms.

**Limitations of Existing Crowdsourcing Platforms:**We observe that commonly available crowdsourcing platforms have limitations in executing our KM scenario. First, they inherently require communities as most of the KM tasks do. In our case, graders (faculty and alumni) assess students' artifacts and both required a certain implicit knowledge to engage, interact and perform within the KM process. Second, the different tasks within our KM process of assessment need coordination and supervision of activities like contributions, verifications and validation in a repetitive way. This is not possible with current architecture of crowdsourcing platforms. Third, training of graders on CLOs specific checking of answer books while paying attention to the alignment perspective poses the biggest limitation, when crowdsourcing platforms like Amazon Mechanical Task and others are considered. Furthermore, quality of evaluations also suffers as a result of random or cost based recruitment of graders. Finally, they should have some form of sociability within them to exercise the mechanisms of reaching an agreement using collaboration and conflict resolution. This mandates the assumption that both faculty and alumni are communities of practice. Currently, only few research-based platforms [31,34,35] offer the concept of teams.

**TABLE 1. TASKS IN OUR EXAMCHECK KM ACTIVITY**

| Tasks | Communities |
|---|---|
| Registration of participant | Student, Faculty, Alumni |
| T-1: Taking photos of exam answer books and uploading them to Web App. | Students |
| T-2: Student Feedback on their attempted exam before checking | Students |
| T-3: Setting distribution criteria for crowdsourcing checking of exam answer books. | Teacher/TAs |
| T-4: Checking of exam answersbooks using community-based crowdsourcing | Faculty and Alumni |
| T-5: Student Feedback after seeing their checked exam answer books | Students |
| T-6: Lesson Learned | Teacher, Faculty, Alumni |

**Opportunities for Community-based Crowdsourcing:**
Community-based crowdsourcing [32,35-37] has been reported, taking advantage of peculiar capabilities and characteristics of crowd worker communities. The linkage of communities to our KM scenario seems obvious; however, a few particular opportunities have kept our motivation thriving:

(1)     Crowdsourcing to communities instead of individual workers makes the collective wisdom of the network through its pro-active members. This not only means improved task quality but also good prospects of getting additional workers.

(2)     Tasks sourced to communities get additional attention during physical interaction between members and not just the workers executing the tasks.

(3)     Communities support redundant execution of tasks and, therefore, more robust to delays and failures.

(4)     Community brings in innovative methods, procedures, and diverse use cases to enrich tasks that require solution related to policy making, conflict resolution, and complex technical and non-technical problems.

(5)     Communities are formed to solve emerging problems, which then collaborative together in opinion forming for knowledge-based outcomes.

(6)     The consensus or knowledge outcome, more specifically, developed by a previous community to a similar problem would be utilized by further communities encountering that same or related problem. These benefits make community-based crowdsourcing fast, economical and having improved quality. It also results in mobilization of knowledge within communities of workers utilizing their interstitial time. Therefore, in our case, proactive graders become more interactive by sharing their expertise and artifacts and make cross-community linkages, resulting in diversification of expertise, and motivation of other members.

# 4.     IMPLEMENTATION AND EXPERIMENTAL SETUP

We have implemented our experimental KM scenario in form of a CS capstone project consisting of a web application along with a built-in crowdsourcing component. However, our plan is to implement a separate community-based crowdsourcing component, called CrowdCheck, as future work. The first author (as supervisor) along with a team of three CS senior students have developed and hosted, a PHP web application on an AWS (Amazon Web Services) based LAMP platform. We opted to save answer book images on Amazon Simple Storage Service (S3) catering for future scalability.

We have tested ExamCheck, on the scenario presented in Section II, within our university. Recruitment of 127 students of EE 213 COAL (Computer Organization and Assembly Language) course form the student community. We registered four CS faculty and few senior students as alumni. We used COAL midterm exam as a test assessment for our experiment. COAL syllabus and all CLOs were added in ExamCheck. The course teacher first collected students' answer books from the examination department. She then created a new assessment in ExamCheck, and for each question: uploaded a sample solution and tagged all related CLOs. At that stage, ExamCheck sent notifications to all registered students starting the KM activity.

The task of uploading images (approximately 625 images) of each answer book solution is crowdsourced to students themselves under supervision of a TA. This relieves teacher/TA/academic staff of manual labor and provide inherent cloud backup of the assessment in question. Students performing this task (an interaction between physical artifact - answer book) identify shortcomings their knowledge about various CLOs. As a knowledge externalization exercise, ExamCheck also allows students to input all shortcoming in their solutions(self-assessment), and use their cognition and memory to recall and tag CLOs with their solution. This tagging and feedback together form yet another online assessment to identify gaps in understanding CLOs as applied to practical problems, however, this is done before grading the answer sheets.

ExamCheck allows the teacher to choose the proportion of the assessments to be distributed in the communities and the number of redundant question for checking in the system. At this point, the theVAB (Virtual Answer Book) generation algorithm executes and generate a task for the grading communities. Each register grader gets an intimation by email (or through web portal page) about the grading task. Graders are allowed to decline the task; however, an optimal threshold is maintained keeping a minimum number of graders in ExamCheck system.

The teacher can forward the assessment by identifying specific communities as graders. The grader can view student assigned CLOs, suggest additional CLOs and mark the current CLOs as wrong. A grader assigns marks on a particular question from the range specified by the teacher. Currently, ExamChecksupports one VAB per grader. ExamCheck maintains a grading progress bar on the teacher portal web page. Once all grading task are completed, the teacher can view the graders' responses, assigned marks and publishes the results.

## 5. FINDINGS AND CHALLENGES

We find that interactions and communications centered around information artifacts (answer sheets in our case) have the potential to attract diverse crowds, forming an interaction space, that externalizes their knowledge (as human computations) for varied reasons (competition, gaming, altruism, status payoffs, or monetary [29]). In this space, human-human and human-artifact interactions promote sharing, transfer and use of existing knowledge, which results in reorganization, evaluations, and validation to create new knowledge. We see our notion of interaction space in the context of the SECI model of knowledge creation by Nonaka [38] and the concept of Ba by Nonaka and Konno [14] and argue that community-based crowdsourcing of human computations is a scalable way to the externalization of human knowledge based on human cognition and memory.

Crowdsourcing to communities instead of individual workers has benefits in knowledge creation and management. Our second finding reveals that self-registration into a community allow workers with an initial set of specific skills. These skills can be refined later based on worker's tasks related performance and interactions within the community. In our opinion, community crowds have many characteristics that can alleviate many existing challenges faced by crowdsourcing and HC research. For example, first we observe that the community building in our experiment is independent of any prior social bonding i.e. worker previously unknown to each other now form a community based on semantic of KM process. Secondly, opposing point-of-views within the KM process context create a kind of social tension, which is healthy for knowledge creation. Thirdly, it promotes externalization of knowledge by generating meta-contents in the form of chats, tagging, categorization, etc. Finally, the emerging consensus or disagreements in human-human and human-artifact interactions are key to knowledge creation and management. The citizen community crowds taking part in this activity become an integrated social group – a community of practice [30], ready to solve similar problems.

In our third finding, we argue that strategically designed HC, performed within the context of a large activity (e.g. KM process), serve as a learning and training platform. In this way, human computations increase implicit knowledge of crowds when they see the whole while focusing on the individual task. For example, in our scenario, first, at the time of uploading the images of their answer books, the student also tag each question with appropriate CLOs and give feedback in the form of self-disclosure about mistakes in their answers. On a similar note, graders while grading the answer sheets, make a mental model of missing CLOs, complex questions, polarized answers, etc. Therefore, the use of learning and training using HC becomes a hands-on approach - like an apprenticeship, by which community can expand itself by integrating and naturalizing new workers. This supervised learning as a result of performing human computational related to a KM process is afforded to crowds as a side effect similar to Luis Von Ahn work on reCAPTCHA [40]. This has a similar duality but a different objective i.e. to utilize wasted human processing power to solve problems that computer cannot solve.

Finally, we find that capturing rich meta-data from various interactions and task executions not only allows efficient execution of future HC tasks but can be used to dynamically change KM process workflow under machine control. Furthermore, different algorithms are allowed to make independent machine interpretation in parallel using this meta-data. In this approach, knowledge is created, as a human-centric way is refined further as a human-machine activity. Capturing of meta-data has been reported by Brambilla et. al. [32] in the form of a control-mart for dynamic adaptations during task executions.

**Challenges:** We foresee four challenges in proposing a framework for knowledge creation and management in citizen community crowds as follows: (C1) Creation of interactions spaces based on artifacts of interest that attract citizen crowds. (C2) The design of a KM process where human-centric tasks externalize community knowledge. (C3) A model to exploit individual worker capabilities in citizen crowds. (C4) A model for aggregation of opinions in citizen crowds.

# 6.    OUR KNOWLEDGE MANAGEMENT FRAMEWORK

This section presents a basic KM model based on community crowds. Most features are based on our observations/findings during the ExamCheck experiment.

**Formation of Interaction Spaces:** Our model (Fig. 2) relies on goals of interest to the community crowds to build virtual interaction spaces using semantic web technologies. Our experiment takes the goal of aligning teaching and assessing with course learning objectives, however, other ways to form interaction spaces could be Q&A (like Stack Overflow and Quora), videos (like Facebook and YouTube), blogs, etc. This is similar to topics based formal interactions in offices, undertaken privately or publicly, and distributed within one or more physical spaces. We expect our interaction spaces to support open interactions and collaboration in citizen community crowds with necessary machine based augmentation. The goal is to attract individual talent, expertise, skills, and memory to support and make use of pluralism and polarization in terms of conflicting opinions to enrich creating knowledge.

**Transition to Knowledge Spaces**: Open interaction spaces lack a common well-defined goal to set the direction and efforts of crowds. Our model specifies goals in terms of KM process (KM-P) containing a workflow of tasks to be performed by human and machines. The novelty of our model is how human-centric tasks are performed by using HC.

Our knowledge space contains five components: (1) describing the KM process in terms of a human-machine workflow, (2) workers for doing the HC using their cognition and memory, (3) worker collaboration patterns specific to KM workflow meaning how they work together (process knowledge = KM workflow + collaboration pattern), (4) what they access in the context of a given KM activity? (5) how machine-processes (algorithms) helps in augmented HC.

Our experiment reveals that interactions spaces executing KM-P tasks affords enabling conditions and necessary context for knowledge exchange and becomes a knowledge space. This shared space, physical or virtual, where humans have shared goals but different vision and approaches to find a solution are ideal for creating, sharing and putting to use knowledge for solving a problem or attaining an innovative outcome [7]. We argue that in knowledge spaces, crowds first expand the problem, due to the availability of a large number of use cases, into everybody's known problem e.g. Q&A done in Stack Overflow/Quora forums. They collectively learn different aspects of the problem by asking related questions, validate the intentions of the requestor, and establish context and other related attributes of the problem, which indirectly help in selecting the best solution for the problem.

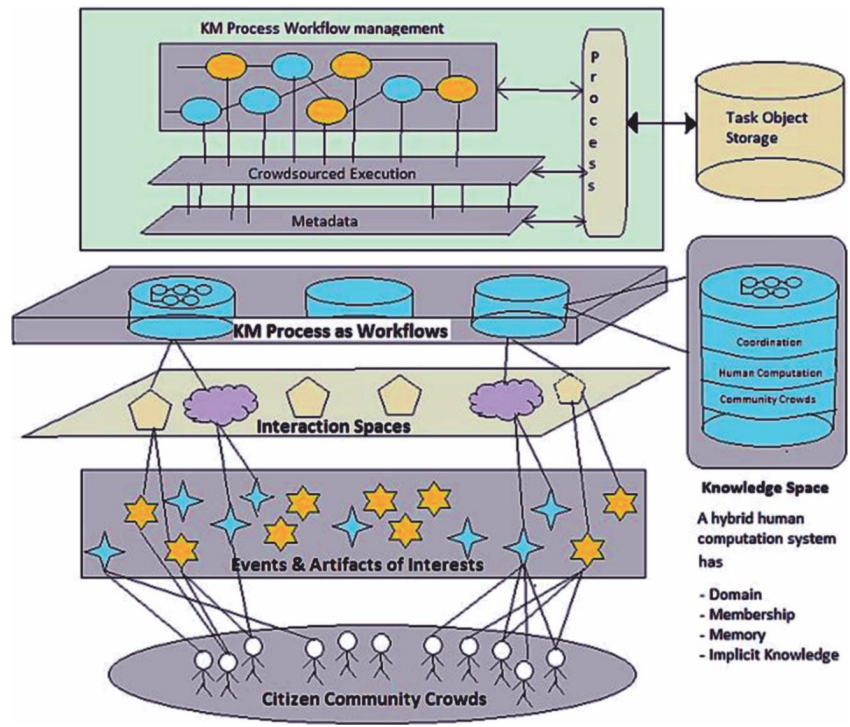

*FIG. 2. KM FRAMEWORK FOR CITIZEN COMMUNITY CROWDS*

A knowledge space may need more than one interaction space to execute its KM-P. This opens up space for machine based interventions and optimizations in forming knowledge spaces. Distributed knowledge spaces may exist as teams develop their own understanding of the knowledge process. A positive social conflict and tension between teams only enrich the context and provide more conditions towards reaching a consensus. These collections of distributed knowledge spaces are places of knowledge sharing, knowledge seeking, cleansing, contextualization and validation of input knowledge.

In knowledge spaces, continuous engagement of participants is accomplished by keeping them informed about their favorite topics, contents, people and tasks. Our model attaches an internal process, called "knowing", to each knowledge space. Knowing process, in its simplest form, send a notification to crowds in different interaction spaces about events: "who knows what", "where is what" and "who has what". We will further improve this concept in our future research. Participation is further ensured by, for example, process meta-data gathered in KS such that crowd contributions accumulate a credit score, which might discount by local, provincial and federal taxes on a yearly basis. The makes crowd contributions as knowledge assets and their interactions are officially recognized as services. KM-P has different tasks, which a knowledge space tries to execute. We plan to execute complex tasks using collaborative teams similar to [3] within the knowledge space. The quality of tasks can be improved using iterative execution reported in [24]. Similarly, redundant execution of the same task improves assured completion mitigating the volunteer nature of crowds.

**The Role of Crowd Worker**: Human-centric tasks are of two natures: (1) those that required human cognition and memory to decide or recall in terms of time, space and context/semantics, and (2) those that requires a human referral (routing) i.e. pointing other human or machine to important resources. These requirements are translated in our model by treating an individual crowd worker as (1) a HCU (Hybrid Compute Unit) for machine and human computations, (2) a HSRU (Human Storage and Recall Unit) where use cases, context, contents, etc. and their linkage graph are stored, (3) a HMSU (Human-Machine Sensing Unit) having hardware based extended sensing capabilities.

These workers form HCC (Human Compute Clouds) for knowledge creation and management utilizing both crowdsourcing and social networking platforms. These HCC can be configured into cloudlets to form smaller units that can be configured into knowledge spaces dynamically and in a distributed fashion to represent different contexts. The characteristics of HCC, cloudlets and networks vary widely based on various conditions e.g. (1) worker availability, (2) their expertise, experience and skills, (3) contents availability including sensor data that is semantically relevant and (4) the networking topology, opportunities for knowledge exchanges and affordance extended by the interaction environments (virtual communities) for KM.

# 7. CONCLUSIONS

This paper has presented a KM scenario and its simple implementation to argue that KM activities can be implemented using strategically designed HC performed on community-based crowdsourcing platforms.

Our experiment ExamCheck, provide the first evidence for our argument and facilitate the exchange of expertise, sharing and transferring knowledge, and giving basic decision support to faculty and students about achieving best learning outcomes in a given undergraduate course. It further identifies that:

(i) Evaluation of information artifacts (e.g. videos, images, blogs etc.) has the potential to become a community-based platform for crowdsourced HC.

(ii) Community crowds have similarities with established KM notions of SECI model and Ba.

(iii) Strategically designed HC; performed within the context of a large activity (e.g. KM process) can serve as a learning and training platform.

(iv) Meta-data captured during execution of HC can be reused later to automate subtasks generation in the context of the same process.

We have presented a basic KM framework for citizen crowds to mitigate challenges observed during our study. Our framework, to the best of our knowledge, utilize a novel knowledge-centric view of the citizen community crowds. The hybrid human-machine computations afford human cognition and memory for performing KM activities as a service.

# 8. FUTURE WORK

Our simple experiment shows that traditional KM should now be seen in a new context of pervasive computing: by bringing people, personal computing devices, IoTs (Internet of Things), and community problem solving together using crowdsourcing HC. We see this transformation of KM into knowledge-centric crowd work as essential requirements for civic innovations, problem-solving and citizen science in future smart cities and societies, where citizens collaborate to perform diverse real-time KM. We will improve our basic KM framework in our future research.

# ACKNOWLEDGEMENTS

Authors thank the reviewers for their valuable comments, Dr. Muhammad Nauman Durrani, for participating in various discussions on different aspect of this paper, and students for their help while prototyping the web application.